\documentclass{elsart}
\usepackage{epsfig}
    \newcommand{\ba}{\begin{eqnarray}}
    \newcommand{\ea}{\end{eqnarray}}
    \newcommand{\be}{\begin{equation}}
    \newcommand{\ee}{\end{equation}}

    \newcommand{\AmS}{{\protect\the\textfont2%
  A\kern-.1667em\lower.5ex\hbox{M}\kern-.125emS}}

\begin{document}
\runauthor{PKU}
\begin{frontmatter}

\title{Lattice study on kaon pion scattering length%
      \ in the $I=3/2$ channel}

\author[PKU]{Chuan Miao},
\author[PKU]{Xining Du},
\author[PKU]{Guangwei Meng}
\author[PKU]{and Chuan Liu}
\address[PKU]{School of Physics\\
              Peking University\\
              Beijing, 100871, P.~R.~China}

\begin{abstract}
Using the tadpole improved Wilson quark action on small, coarse
and anisotropic lattices, $K\pi$ scattering length in the $I=3/2$
channel is calculated within quenched approximation. The results
are extrapolated towards the chiral and physical kaon mass region.
Finite volume and finite lattice spacing errors are also analyzed
and a result in the infinite volume and continuum limit is
obtained. Our result is compared with the results obtained using
Roy equations, Chiral Perturbation Theory, dispersion relations
and the experimental data.
\end{abstract}
\end{frontmatter}


\section{Introduction}

 Low-energy $K\pi$  scattering experiment is an
 important experimental method in the study of interactions among
 mesons~\cite{matison74:Kpi_exp_a,johannesson73:Kpi_exp_b,shaw80:Kpi_exp_c}.
 It is also a good testing ground for our understanding
 of the low-energy structure of Quantum Chromodynamics (QCD).
 Chiral Perturbation Theory~\cite{meissner91:Kpi_a,meissner91:Kpi_b},
 Roy equations~\cite{buettiker03:Kpi_Roy},
 dispersion relations~\cite{lang77:Kpi_dispersion,zheng04:Kpi_dispersion}
 and other theoretical methods~\cite{buettiker01:Kpi_sumrule} have
 been used in the study of low-energy $K\pi$ scattering.
 However, if the hadrons being scattered involve strange quarks,
 as in the case of $K\pi$ scattering, predictions
 within Chiral perturbation theory usually suffer from sizable
 corrections due to $SU(3)$ flavor symmetry breaking, as compared with
 the case of pion-pion scattering.
 Lattice QCD is a genuine non-perturbative method which
 can handle hadron-hadron scattering at low-energies.
 Therefore, a lattice QCD calculation will offer an important
 and independent check on these results.
 In fact, pion-pion~\cite{gupta93:scat,fukugita95:scat,jlqcd99:scat,%
 chuan02:pipiI2,JLQCD02:pipi_length,juge03:pipi_length,%
 ishizuka03:pipi_length,CPPACS03:pipi_phase,CPPACS03:pipi_phase_unquench},
 pion-nucleon~\cite{fukugita95:scat} and
 kaon-nucleon~\cite{chuan04:KN} scattering
 have been studied using lattice QCD.
 In this letter, we report our quenched lattice
 results on the $K\pi$ scattering length in the $I=3/2$ channel.
 This quantity is of phenomenological importance for
 $K\pi$ scattering since the value of the scattering
 length in the $I=3/2$ channel is correlated with
 that in the $I=1/2$ channel and the value of the
 latter is crucial for the determination of
 the possible $\kappa$ resonance in the $I=1/2$
 channel~\cite{zheng04:Kpi_dispersion}.

 The lattice gauge action that is used in our study is
 the tadpole improved clover action on anisotropic
 lattices~\cite{colin97,colin99}.
 Using this gauge action, glueball and hadron spectra have been studied
 within quenched approximation \cite{colin97,colin99,%
 chuan01:gluea,chuan01:glueb,chuan01:canton1,chuan01:canton2,chuan01:india}.
 In our previous works, configurations generated from this
 improved action have also been utilized to calculate the
 $\pi\pi$ scattering lengths in the $I=2$ channel
 \cite{chuan02:pipiI2} and the $KN$ scattering length in
 the $I=1$ channel~\cite{chuan04:KN}.
 The fermion action used in this study is the
 tadpole improved clover Wilson action on anisotropic
 lattices \cite{klassen99:aniso_wilson,chuan01:tune,chuan02:pipiI2}:
 \ba
 \label{eq:fmatrix}
 {\mathcal M}_{xy} &=&\delta_{xy}\sigma + {\mathcal A}_{xy}
 \nonumber \\
 {\mathcal A}_{xy} &=&\delta_{xy}\left[{1\over 2\kappa_{\max}}
 +\rho_t \sum^3_{i=1} \sigma_{0i} {\mathcal F}_{0i}
 +\rho_s (\sigma_{12}{\mathcal F}_{12} +\sigma_{23}{\mathcal F}_{23}
 +\sigma_{31}{\mathcal F}_{31})\right]
 \nonumber \\
 &-&\sum_{\mu} \eta_{\mu} \left[
 (1-\gamma_\mu) U_\mu(x) \delta_{x+\mu,y}
 +(1+\gamma_\mu) U^\dagger_\mu(x-\mu) \delta_{x-\mu,y}\right] \;\;,
 \ea
 where various coefficients in the fermion matrix
 ${\mathcal M}$ are given by:
 \ba
 \eta_i &=&\nu/(2u_s) \;\;, \;\;
 \eta_0=\xi/2 \;\;, \;\;\sigma=1/(2\kappa)-1/(1\kappa_{max})\;\;,
 \nonumber \\
 \rho_t &=& c_{SW}(1+\xi)/(4u^2_s) \;\;, \;\;
 \rho_s = c_{SW}/(2u^4_s) \;\;.
 \ea
 Among the parameters which appear in the fermion matrix, $c_{SW}=1$
 is the tree-level clover coefficient and $\nu$ is the so-called
 bare velocity of light, which has to be tuned non-perturbatively
 using the single pion dispersion relations \cite{chuan01:tune}.
 In the fermion matrix~(\ref{eq:fmatrix}), the bare quark mass
 dependence is singled out into the parameter $\sigma$ and the
 matrix ${\mathcal A}$ remains unchanged if the bare quark mass is
 varied.  This shifted structure of the matrix ${\mathcal M}$ can be utilized
 to solve for quark propagators at various values of valance quark mass
 $m_0$ (or equivalently $\kappa$)
 at the cost of solving only the lightest valance
 quark mass value at $\kappa=\kappa_{\max}$,
 using the so-called  Multi-mass Minimal
 Residual ($M^3R$ for short) algorithm %
 \cite{frommer95:multimass,glaessner96:multimass,beat96:multimass}.
 This is particularly advantageous in a quenched calculation
 since one needs the results at various quark mass values
 to perform the chiral extrapolation.

 The basic procedure for the calculation of $K\pi$ scattering
 length is similar to that adopted in the $\pi\pi$ scattering length
 calculation~\cite{chuan02:pipiI2}.

 \section{Numerical calculation of the scattering length}
 \label{sec:theory}

 In order to calculate the elastic scattering lengths
 for hadron-hadron scattering on the lattice,
 or the scattering phase shifts in general, one uses L\"uscher's
 formula which relates the exact energy level of two hadron states
 in a finite box to the elastic scattering phase shift in the continuum.
 In the case of $K\pi$ scattering at
 zero relative three momentum, this formula amounts to a relation
 between the {\em exact} energy $E^{(I)}_{K\pi}$ of the
 $K\pi$ system with vanishing relative momentum
 in a finite box of size $L$ with isospin $I$, and
 the corresponding scattering length
 $a^{(I)}_0$ in the continuum.
 This formula reads \cite{luscher86:finiteb}:
 \be
 \label{eq:luescher}
 E^{(I)}_{K\pi}-(m_K+m_\pi)=-\frac{2\pi a^{(I)}_0}{\mu_{K\pi} L^3}
 \left[1+c_1\frac{a^{(I)}_0}{L}+c_2(\frac{a^{(I)}_0}{L})^2
 \right]+O(L^{-6}) \;\;,
 \ee
 where $c_1=-2.837297$, $c_2=6.375183$ are numerical constants
 and $\mu_{K\pi}=m_Km_\pi/(m_K+m_\pi)$ is the reduced mass of the
 $K\pi$ system. In this letter, the $K\pi$ scattering length $a^{(3/2)}_0$
 in the isospin $I=3/2$ channel will be calculated.

 To measure the hadron mass values $m_\pi$, $m_K$, $m_\rho$ and
 to extract the energy shift $\delta E^{(3/2)}_{K\pi}$,
 one constructs the correlation
 functions from the corresponding one meson and two meson operators
 in the appropriate symmetry channel. For example,
 the $K\pi$ operator in the $I=3/2$ channel is given by:
 \be
 O^{I=3/2}_{K\pi}(t) = K^+_0(t)\pi^+_0(t+1)\;\;,
 \ee
 where $K^+_0(t)$ and $\pi^+_0(t)$ are the zero momentum kaon and
 pion operators respectively. The $K\pi$ correlation
 function in the $I=3/2$ channel then reads:
 \be
 C^{I=1}_{K\pi}(t)=\langle
 O^{I=3/2\dagger}_{K\pi}(t)
 O^{I=3/2}_{K\pi}(0)\rangle \;\;.
 \ee
 Numerically, it is more advantageous to construct
 the ratio of the correlation functions defined above:
 \be
 {\mathcal R}^{I=1}(t) = C^{I=3/2}_{K\pi}(t) / (C_K(t)C_\pi(t)) \;\;.
 \ee
 In this case, one uses the linear fitting function:
 \be
 \label{eq:linear_fit}
 {\mathcal R}^{I=3/2}(t) \stackrel{T >> t >>1}{\sim} 1-\delta E^{(3/2)}_{K\pi}t
 \;\;,
 \ee
 to determine the energy shift $\delta E^{(3/2)}_{K\pi}$.
 $K\pi$ correlation function, or equivalently, the ratio
 ${\mathcal R}^{I=3/2}(t)$ is constructed from quark propagators,
 which are obtained using the Multi-mass Minimal
 Residue algorithm with wall sources.
 Periodic boundary condition is applied to all three
 spatial directions while in the temporal direction, Dirichlet
 boundary condition is utilized.

 Configurations used in this calculation are generated using the pure
 gauge action for $6^340$, $8^340$ and $10^350$ lattices with the gauge
 coupling $\beta=1.9$, $2.2$, $2.4$, $2.6$ and $3.0$. The spatial lattice
 spacing $a_s$ is roughly between $0.1$fm and $0.4$fm while
 the spatial physical size of the lattice ranges from
 $0.7$fm to $4.0$fm. For each set of parameters, several hundred
 decorrelated gauge field configurations are used to measure
 the fermionic quantities. Statistical errors are all analyzed
 using the usual jack-knife method.
 Single pion, kaon and rho  mass values
 are obtained from the plateau of their corresponding effective
 mass plots. The fitting interval is automatically
 chosen by minimal $\chi^2$ per degree of freedom.
 Due to the usage of finer lattice spacing in
 the temporal direction, good plateau behavior was observed
 in these effective mass plots. Therefore, contaminations from
 excited states should be negligible.
 These mass values will be utilized in the chiral
 extrapolation.
%

 After obtaining the energy shifts $\delta E^{(3/2)}_{K\pi}$,
 these values are substituted into L\"uscher's formula to solve
 for the scattering length $a^{(3/2)}_0$ for all possible
 hopping parameter pairs: $(\kappa_u,\kappa_s)$, which
 corresponds to the up (down) and the strange
 bare quark mass values, respectively.
 This is done for lattices of all sizes being
 simulated and for all values of $\beta$.
 From these results, attempts are made to perform
 an extrapolation towards the chiral, infinite volume
 and continuum limit.

 \section{Extrapolations of the scattering length}
 \label{sec:extrap}

 The chiral extrapolations of physical quantities involving both
 the up (down) and the strange quarks consists of two
 steps. In the first step, the bare strange quark mass, or equivalently
 the corresponding hopping parameter $\kappa_s$, is kept
 fixed while the hopping parameter of the up/down quark, $\kappa_u$,
 is brought to their physical value $\kappa^{(phy)}_u$.
 The precise value of $\kappa^{(phy)}_u$ can be obtained by
 inspecting the chiral behavior of the pseudo-scalar (pion)
 and the vector meson (rho) mass values.
 In the second step, one fixes the up/down quark
 hopping parameter at its physical value obtained in the first step
 and extrapolate/interpolate in
 the strange quark mass. The physical strange quark hopping parameter
 $\kappa^{(phy)}_s$ is determined by demanding the mass of
 the kaon being exactly at its physical value (the so-called K-input).

 We now come to the chiral extrapolation of the scattering length.
 In the chiral limit, the $K\pi$ scattering length in
 the $I=3/2$ channel is given by the current algebra result:
 \be
 \label{eq:CA}
 a^{(3/2)}_0 =-{1 \over 8 \pi} {\mu_{K\pi} \over f^2_\pi} \;\;,
 \ee
 where $a^{(3/2)}_0$ is the $K\pi$ scattering length in
 the $I=3/2$ channel and $f_\pi\sim 93$MeV is the pion
 decay constant.
 To perform the chiral extrapolation of the scattering length,
 it is more convenient to use the quantity
 $F=a^{(3/2)}_0m^2_\rho/\mu_{K\pi}$, which in the chiral
 limit reads: \cite{chuan02:pipiI2}
 \be
 \label{eq:chiral}
 F\equiv {a^{(3/2)}_0m^2_\rho \over \mu_{K\pi}}
 =-{1 \over 8 \pi} {m^2_\rho \over f^2_\pi}
 \sim  -2.728\;\;,
 \ee
 where the final numerical value is obtained
 by substituting in the experimental values for
 $m_\rho\sim 770$MeV and $f_\pi\sim 93$MeV.
 One-loop corrections reduce the central value
 of the scattering length by
 about $28$\%, with
 an estimated error of $40$\% \cite{meissner91:Kpi_a,meissner91:Kpi_b}.
 The factor $F$ can be calculated on the lattice with good
 precision {\em without} the lattice calculation of meson
 decay constants.
 Since we have calculated the factor $F$ for several different
 values of valance quark mass, we could make a chiral
 extrapolation and extract the corresponding results in the chiral limit.

\begin{figure}[thb]
 \begin{center}
  \vspace{-50mm}
 \includegraphics[height=10.0cm,angle=0]{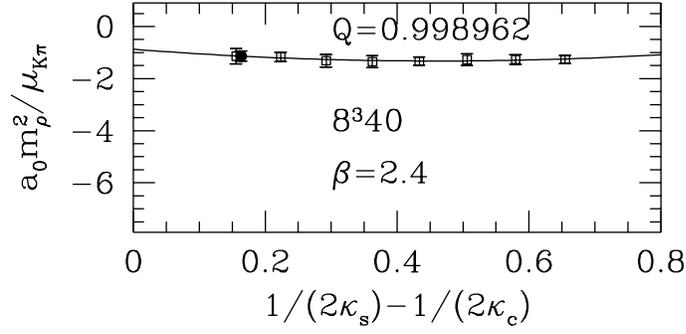}
 \end{center}
 \caption{Chiral extrapolation for the quantity
 $F\equiv a^{(1)}_0m^2_\rho/\mu_{KN}$ for our simulation results
 at $\beta=2.4$ on $8^340$ lattices. In this plot,
 quantity $F$, with the up and down quark mass already
 extrapolated to the chiral limit, is plotted (open squares)
 as a function of strange quark mass parameter $1/(2\kappa_s)$.
 The line represents the quadratic
 interpolation/extrapolation to the physical strange quark mass, where
 the result is depicted as a solid square.
 \label{fig:chiral_extrapolation}}
 \end{figure}

 In chiral extrapolations, we have adopted a
 quadratic functional form and
 the fitting range of the extrapolation is self-adjusted by the
 program to yield a minimal $\chi^2$ per degree of freedom.
 First, we extrapolate the factor $F$ in the up/down quark mass values.
 Second, the extrapolated values for the factor $F$ are then
 extrapolate/interpolate in the strange quark mass.
 In Fig.~\ref{fig:chiral_extrapolation}, we have shown the
 extrapolation in the strange quark mass
 for one of our simulation points.
 The resulting factor $F$ after the up and down
 quark mass extrapolation are plotted in
 Fig.~\ref{fig:chiral_extrapolation} versus
 the strange quark mass parameter $1/(2\kappa_s)$.
 The line represents the quadratic  extrapolation/interpolation
 and the final result is depicted as a solid square at the physical strange
 quark mass.  The fitting quality $Q$ for this fit is also displayed.

\begin{figure}[thb]
 \begin{center}
   \vspace{-10mm}
 \includegraphics[height=10.0cm,angle=0]{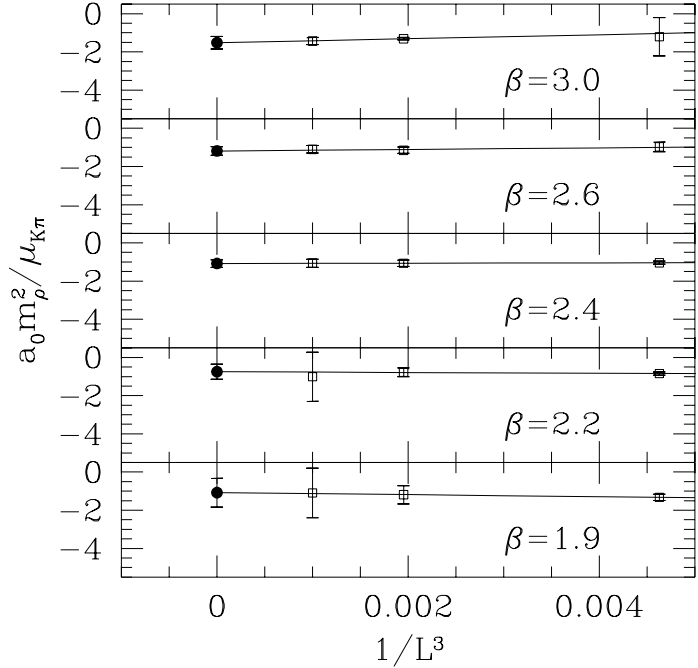}
 \end{center}
 \caption{Infinite volume extrapolation for the quantity
 $F=a^{(3/2)}_0m^2_\rho/\mu_{K\pi}$ obtained from our simulation results
 at $\beta=3.0$, $2.6$, $2.4$, $2.2$ and $1.9$.
 The straight line represents the linear extrapolation in $a_s/r_0$.
 The extrapolated results are shown with solid squares at $L^{-3}=0$.
 \label{fig:volumeX}}
 \end{figure}

 After the chiral extrapolation, we now turn to study
 the finite volume effects of the simulation. According
 to formula~(\ref{eq:luescher}), the quantity $F$ obtained
 from finite lattices differs from
 its infinite volume value by corrections of the form $1/L^3$.
 In Fig.~\ref{fig:volumeX},
 we have shown the infinite volume extrapolation for
 our  simulation points at
 $\beta=3.0$, $2.6$, $2.4$, $2.2$ and $1.9$.
 The extrapolated results for the factor $F$ are
 then used for the continuum limit extrapolation.

 \begin{figure}[thb]
 \begin{center}
  \vspace{-50mm}
 \includegraphics[height=10.0cm,angle=0]{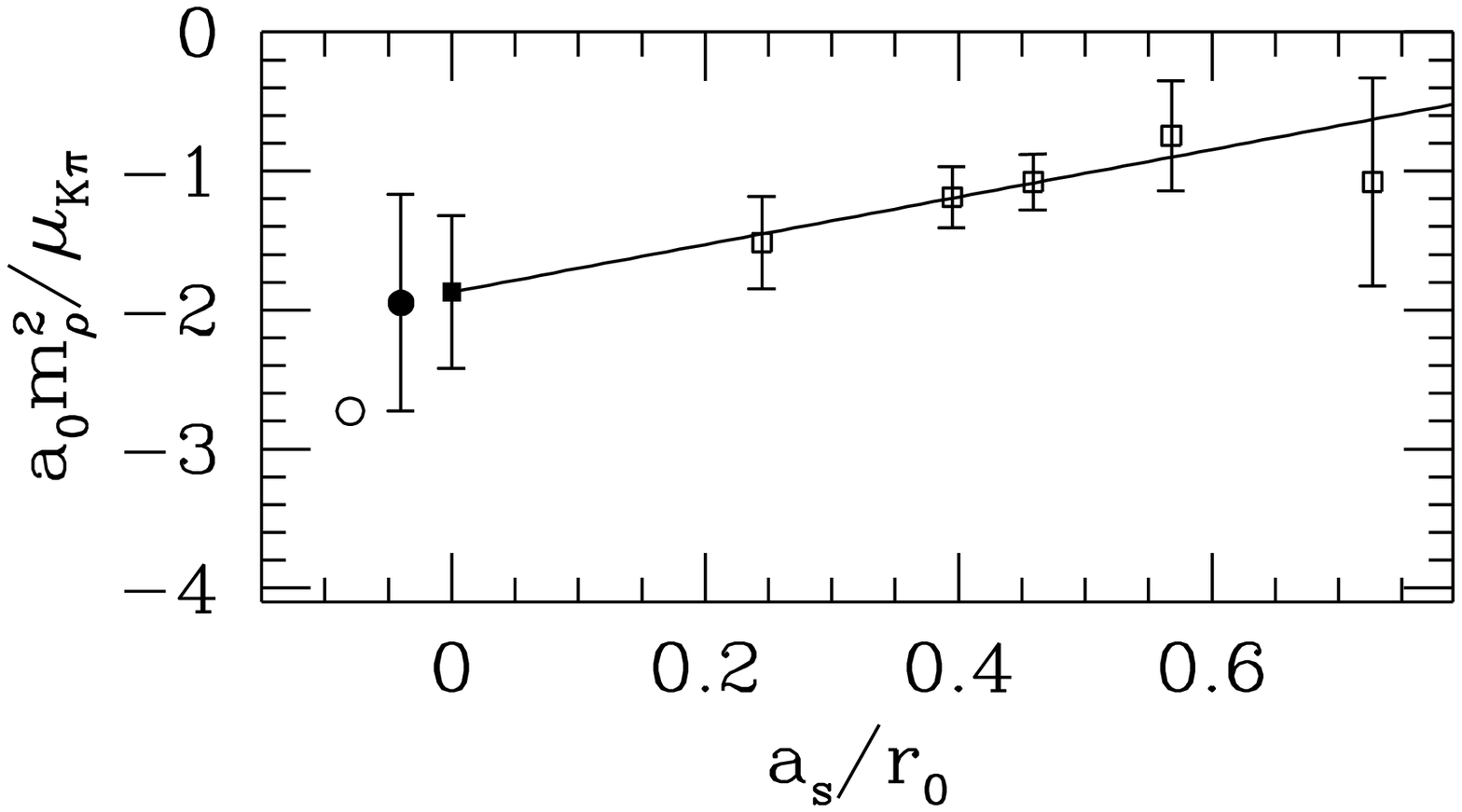}
 \end{center}
 \caption{Continuum extrapolation for the quantity
 $F=a^{(3/2)}_0m^2_\rho/\mu_{K\pi}$ obtained from our simulation results
 at $\beta=3.0$, $2.6$, $2.4$, $2.2$ and $1.9$.
 The straight line represents the linear extrapolation in $a_s/r_0$.
 The extrapolated result is also shown as solid square with
 the corresponding error. Also shown are the
 result from current algebra and dispersion relations~\cite{lang77:Kpi_dispersion}.
 They coincide numerically and the value is show as an open circle.
 The chiral perturbation theory result is shown as
 a solid circle with error bar.
 \label{fig:continuum_extrapolation}}
 \end{figure}
 Finally, we can make an extrapolation towards the
 continuum limit by eliminating the finite lattice
 spacing errors. Since we have used the tadpole
 improved clover Wilson action, all physical quantities
 differ from their continuum counterparts by
 terms that are proportional to $a_s$. The physical
 value of $a_s$ for each value of $\beta$ can be
 found from Ref. \cite{colin99,chuan01:india}.

 The result of the continuum extrapolation is shown in
 Fig.~\ref{fig:continuum_extrapolation} where
 the results from the chiral and infinite volume
 extrapolation discussed above are indicated as
 data points in the plot for all $5$ values of
 $\beta$ that have been simulated.
 The straight line shows
 the extrapolation towards the $a_s=0$ limit and the
 extrapolated result is also shown as a solid square together with
 the chiral result from Ref.\cite{meissner91:Kpi_a,meissner91:Kpi_b}
 which is shown as the open (tree-level, or current algebra result)
 and filled (one-loop) circles. Result from dispersion relations
 from Ref.~\cite{lang77:Kpi_dispersion} happens to coincide with the current
 algebra result (open circle) numerically.
 It is seen that these results are compatible within error bars.
 To summarize, we obtain from the linear extrapolation the following
 result for the quantity
 $F=a^{(3/2)}_0m^2_\rho/\mu_{K\pi}=(-1.87\pm 0.55)$.
 If we substitute in the physical values, we obtain the
 $K\pi$ scattering length in the $I=3/2$ channel:
 $a^{(3/2)}_0m_\pi= (-0.048 \pm 0.014)$, which is to be compared
 with the  one-loop chiral result of $(-0.05 \pm 0.02)$ and
 the tree-level result (and also the early dispersion relation result) $-0.07$.
 Our lattice study also indicates a value consistent with
 the chiral result that is smaller in magnitude than the experimental result.
 The experimental results for the $K\pi$ scattering
 length $a^{(3/2)}_0m_\pi$ show sizable
 variation~\cite{matison74:Kpi_exp_a,johannesson73:Kpi_exp_b,shaw80:Kpi_exp_c}.
 It lies between $-0.13$ and $-0.05$, significantly larger in
 magnitude compared with chiral and early dispersion relation results.
 In a recent analysis~\cite{zheng04:Kpi_dispersion},
 a value of $a^{(3/2)}_0m_\pi=-0.129\pm 0.006$
 was obtained using dispersion relations incorporated
 with chiral perturbation theory to fit the experimental data.

 \section{Conclusions}
 \label{sec:conclude}

 In this paper, we have calculated kaon-pion scattering
 lengths in isospin $I=3/2$ channel using quenched
 lattice QCD. It is shown that such a calculation
 is feasible using relatively small, coarse and anisotropic lattices.
 The calculation is done using the
 tadpole improved clover Wilson action on anisotropic
 lattices. Simulations are performed on lattices
 with various sizes, ranging from $0.7$fm to about
 $4$fm and with five different values of lattice spacing.
 Quark propagators are measured with different
 valence quark mass values. These enable us to
 explore the finite volume and the finite
 lattice spacing errors in a systematic fashion.
 The infinite volume extrapolation is made.
 The lattice result for the scattering length is
 extrapolated towards the chiral
 and continuum limit where a result consistent with
 Chiral Perturbation Theory and early dispersion relation
 calculation is found. Our result for the scattering length
 is smaller in magnitude when compared with the
 experimental data results.

 \section*{Acknowledgments}

 This work is supported by the National Natural
 Science Foundation (NFS) of China under grant
 No. 90103006, No. 10235040 and supported by the Trans-century fund from Chinese
 Ministry of Education. C. Liu would like
 to thank Prof.~H.~Q.~Zheng and Prof.~S.~L.~Zhu for helpful discussions.


\end{document}